\documentclass[12pt]{article}

\usepackage{graphicx}
\usepackage{comment}
\usepackage{amsmath}
\usepackage{amssymb}
\usepackage{fullpage}

\numberwithin{equation}{section}

\includecomment{versiona}

\newlength{\dummysp}
\newcommand{\tr}{\mathop{{\hbox{Tr} \, }}\nolimits}
\newcommand{\sech}{\mathop{{\hbox{sech}  }}\nolimits}
\newcommand{\csch}{\mathop{{\hbox{csch}  }}\nolimits}

\newcommand{\half}{\frac{1}{2}}

\newcommand{\beq}{\begin{eqnarray}}
\newcommand{\eeq}{\end{eqnarray}}
\newcommand{\nnn}{ \nonumber \\ }
\newcommand{\ddd}{ \nnn && }
\newcommand{\p}{{\partial}}
\newcommand{\e}{{\epsilon}}
\newcommand{\s}{{\sigma}}

\newcommand{\ord}[1]{{{\cal O}(#1)}}
\newcommand{\gappeq}{\mathrel{\rlap {\raise.5ex\hbox{$>$}}
{\lower.5ex\hbox{$\sim$}}}}
\newcommand{\lappeq}{\mathrel{\rlap{\raise.5ex\hbox{$<$}}
{\lower.5ex\hbox{$\sim$}}}}
\newcommand{\myref}[1]{(\ref{#1})}

\newcommand{\ben}{\begin{enumerate}}
\newcommand{\een}{\end{enumerate}}
\newcommand{\bit}{\begin{itemize}}
\newcommand{\eit}{\end{itemize}}
\newcommand{\Cbf}{{\bf C}}
\newcommand{\Rbf}{{\bf R}}

\newcommand{\Ncal}{{\cal N}}

\newcommand{\psib}{{\bar \psi}}

\newcommand{\sqtw}{\sqrt{2}}

\newcommand{\Zbf}{{\bf Z}}

\newcommand{\zbar}{{\bar z}}

\newcommand{\ibar}{{\bar \imath}}
\newcommand{\jbar}{{\bar \jmath}}

\newcommand{\chib}{{\bar \chi}}
\newcommand{\Psib}{{\bar \Psi}}

\def\[{\left [}
\def\]{\right ]}
\def\({\left (}
\def\){\right )}

\def\nott#1{\setbox0=\hbox{$#1$}                
   \dimen0=\wd0                                 
   \setbox1=\hbox{/} \dimen1=\wd1               
   \ifdim\dimen0>\dimen1                        
      \rlap{\hbox to \dimen0{\hfil/\hfil}}      
      #1                                        
   \else                                        
      \rlap{\hbox to \dimen1{\hfil$#1$\hfil}}   
      /                                         
   \fi}                                         %

\begin{document}

\begin{titlepage}

\begin{center}
{\bf \large Warped compactifications and holographic duality}
\end{center}

\bigskip

\bigskip

\begin{center}
Joel Giedt \\
{\it Department of Physics, Applied Physics and Astronomy \\
Rensselaer Polytechnic Institute, 110 8th Street, Troy, NY 12180 USA }
\end{center}

\bigskip

\bigskip

\begin{abstract}
We review warped compactifications of superstring theory with some attention
to the limit in which these resemble ``bottom-up'' phenomenological models.
In addition to some discussion of the original Klebanov-Witten and Klebanov-Strassler
set-ups, we also touch on various generalizations of the geometry that have
been considered.
Various other systems with a holographic duality are also briefly reviewed.
The point of this latter exploration is to illustrate how far beyond the
standard $AdS_5 \times S^5$ set-up things have moved over the years.
\end{abstract}

\end{titlepage}


\section{Introduction}
Strongly coupled extensions to the standard model of particle physics have a long
history, beginning with ideas that became known as technicolor \cite{Susskind:1978ms,Weinberg:1979bn}.
Mostly, these ideas are now known as composite Higgs.
Various elaborations on this scenario were made to address various phenomenological
problems with the original proposal. For instance, walking
technicolor was introduced to address the scale of
fermion masses \cite{Holdom:1981rm,
Holdom:1984sk,Yamawaki:1985zg,Bando:1986bg,Appelquist:1986an,
Appelquist:1986tr,Appelquist:1987fc}.  However, the space of possible theories is
vast once extensions are included to address issues of flavor.  See \cite{Appelquist:1993sg}
for just one example.
Furthermore, it has proven difficult to make reliable predictions because
the theories are beyond the reach of perturbation theory.  As a result, various
lattice studies have been conducted in the last several years, using large scale
numerical simulations.  (See for example \cite{Kribs:2016cew,DeGrand:2019vbx}
and references therein.) 
For each theory several years must be devoted to developing
the codes, exploring the parameter space and extracting physics results with
increasing levels of sophistocation.  Even after all of this, there are significant
limitations coming from the difficulty of simulating light states, of signal-to-noise
when fermion-disconnnected diagrams are involved, and the effect of excited states
in the regime where this signal is strongest.  It should also be noted that the
theories that are actually simulated are ones that are very QCD-like, so that
theories of for instance extended technicolor where chiral gauge theories might
play a role are never addressed.

However, an alternative approach to studying strongly
coupled gauge theories has been available since the late 1990s, namely the holographic
correspondence \cite{Maldacena:1997re}. 
Here, a weakly coupled gravitational alternative can be utilized,
albeit in a larger number of spacetime dimensions.  A key component of many of these
dualities is the use of anti-deSitter (AdS) space in the extra-dimensional theory.
In the case of gauge theories with four-dimensional spacetimes (4d), the AdS
space is five-dimensional (AdS$_5$).  When this gravitational theory comes to
us from string theory, there is in addition another five dimensions, and the
complete geometry is AdS$_5$ $\times$ $X_5$, at least in some regime of the
``radial'' coordinate (variously $r$, $z$, $\tau$ or $y$).  Typically $X_5$ is compact.
The original example has $X_5 = S^5$, the five-dimensional sphere.  However,
in the case of Type IIB string theory this leads to a maximally supersymmetric
gauge theory in 4d, $\Ncal=4$ super-Yang-Mills, with no matter and no possibility
for a complex representation.  This leads one to wonder what other choices of $X_5$ 
could be made, how much supersymmetry they would have, and what the dual
gauge theory happens to be.  In this review we discuss some of the investigations
in this direction.  Furthermore, there are modifications to the geometry or
string theory set-up that can lead to interesting matter representations.
The initial efforts in this direction involve adding ``probe branes,'' i.e.,
neglecting back-reaction on the geometry, but still having objects that can
carry ``flavor'' quantum numbers.  

There are a couple of features of the original AdS/CFT formulation that need
modification when it comes to realistic particle phenomenology.  As mentioned, the
Maldacena system has too much supersymmetry.  We need $\Ncal=1$ SUSY in order
to have a chiral gauge theory like the Standard Model.  Second, we have to
move away from a conformal system.  Again, the Standard Model has many
mass scales and is not conformal.  In particular, we need to accommodate
phenomena such as confinement, spontaneous symmetry breaking and chiral condensates.

These matters are addressed to a great extent with compactifications
on the deformed or resolved conifold, which has $T^{1,1}$ for its base
instead of $S^5$.  The five-dimensional manifold $T^{1,1}$ is topologically
equivalent to $S^3 \times S^2$.  In the case of the deformed conifold,
the $S^3$ remains finite at the tip of the cone ($r \to 0$) whereas the
$S^2$ shrinks to zero size.  For the resolved conifold the situation is
reversed.  These correspond to an IR cutoff in the theory.  In the
Klebanov-Strassler construction \cite{Klebanov:2000hb} this is dual to confinement in the
gauge theory, complete with a tower of Seiberg dualities.

In this review we also take the opportunity to briefly discuss
some other versions of holography that are of interest.  One is
holographic cosmology, which makes predictions that differ somewhat
from conventional cosmology \cite{Afshordi:2016dvb}.  It is remarkable that aspects of
cosmology could be described by studying a 3d field theory,
say, using lattice techniques.  We also present some aspects
of the D1-D5 system \cite{deBoer:1998kjm}, and related theories of 6d supergravity
formulated on $AdS_3 \times S^3$.  These give rise to 2d CFTs,
about which a lot can be said due to the large symmetry algebra.

\section{The conifold}
The conifold is described in terms of four complex coordinates satisfying
\beq
z_1 z_2 = z_3 z_4
\label{zeq}
\eeq
or equivalently
\beq
w_1^2 + w_2^2 + w_3^2 + w_4^2 = 0
\label{wsalladdtozero}
\eeq
The transformation between $z_i$ and $w_i$ is straightforward to work out
and has been tabulated many times in the literature.  It is given by
\beq
&& z_1 = w_1 + i w_2, \quad
z_2 = w_1 - i w_2 \nnn
&& z_3 = -w_3 + i w_4, \quad
z_4 = w_3 + i w_4
\eeq
Four complex coordinates subject to one complex constraint is equivalent
to eight real coordinates subject to two real constraints, leaving a
six-dimensional ``surface'' $Y$.  Given the form of the above equations,
which allows for a rescaling of all complex coordinates, $z_i \to e^\lambda z_i$,
 there
is one overall scale, which will be our radial coordinate $r$, and five
compact coordinates, which are angles $\psi, \theta_1, \theta_2, \phi_1, \phi_2$.
The geometry has the structure of a cone, $Y = \Rbf \times X_5$.
Then it is found that the solution to Eq.~\myref{zeq} is
\beq
z_1 &=& r^{3/2} e^{\frac{i}{2} ( \psi - \phi_1 - \phi_2 )}
\sin \frac{\theta_1}{2} \sin \frac{\theta_2}{2}
\nnn
z_2 &=& r^{3/2} e^{\frac{i}{2} ( \psi + \phi_1 + \phi_2 )}
\cos \frac{\theta_1}{2} \cos \frac{\theta_2}{2}
\nnn
z_3 &=& r^{3/2} e^{\frac{i}{2} ( \psi + \phi_1 - \phi_2 )}
\cos \frac{\theta_1}{2} \sin \frac{\theta_2}{2}
\nnn
z_4 &=& r^{3/2} e^{\frac{i}{2} ( \psi - \phi_1 + \phi_2 )}
\sin \frac{\theta_1}{2} \cos \frac{\theta_2}{2}
\label{zsoln}
\eeq

The conifold is a non-compact Calabi-Yau (CY) manifold.  In string
theory, a warping of the spacetime is obtained by placing a
stack of D3-branes at the 
conical singularity $r \to 0$.  In some of the
more advanced models, the singularity may be resolved or deformed to make it non-singular.  In the
gauge theory this translates into infrared (IR) dynamics associated with a cutoff scale.
In terms of the IR dynamics, there is a special role for fluxes through the cycles of the CY,
as will be seen below.  Indeed, this can break conformal invariance and lead to
confinement at low energies.

The compact manifold is described by:
\beq
ds_{T^{1,1}}^2 = \frac{1}{6} \sum_{i=1}^2 ( d\theta_i^2 + \sin^2 \theta_i d \phi_i^2 )
+ \frac{1}{9} ( d \psi + \sum_{i=1}^2 \cos \theta_i d \phi_i )^2
\eeq
This is one of the manifolds found by Romans, related to preserving some
supersymmetry under compactification \cite{Romans:1984an}.

The presence of the D3 branes warps the space.  
So on top of the conifold we have warp factors $H(r)$:
\beq
ds^2 = H(r)^{-1/2} \eta_{\mu\nu} dx^\mu dx^\nu + H(r)^{1/2} ( dr^2 + r^2 ds^2_{T^{1,1}} )
\label{10dmetric}
\eeq
where
\beq
H(r) = 1 + \frac{L^4}{r^4}
\label{KWH}
\eeq
Here, when comparison is made to the string theory, one finds $L^4 = 4 \pi g_s N \alpha'^2$
where $g_s$ is the string coupling and $\alpha'$ is the Regge slope parameter.
It is amusing that this gives a power-law warp factor, whereas in Randall-Sundrum type
discussions one often finds an exponential warp factor.  This is a reflection of the
general coordinate reparameterization invariance of general relativity:  we can make a
change of coordinates which drastically changes the way the space looks.

The geometry \myref{10dmetric} is the basic geometry of the Klebanov-Witten
(KW) construction \cite{Klebanov:1998hh}.  In that case, the dual gauge theory is
superconformal, reflecting the conical singularity, but with reduced supersymmetry,
reflecting the $T^{1,1}$ compact space.

\section{Probe branes}
There are low energy phenomena that we would like to insert into 
these super-Yang-Mills (SYM) theories.
In particular, we would like to have the mesons and baryons of quantum chromodynamics (QCD).
For this, we need fields in the fundamental representation of the color group $SU(N_c)$,
and which also carry flavor quantum numbers under $SU(N_f)$.  In order to accomplish this,
a traditional pathway has been to introduce ``probe'' D7 branes.  The reason why the probe
approximation is used is that the backreaction of D7 branes on the geometry is difficult to
account for, though there have been steps in that direction \cite{Aharony:1998xz}.

Levi and Ouyang have studied the mesons in the 
Klebanov-Witten with D7 probe branes scenario \cite{Levi:2005hh}.
They work from an embedding based on the coordinates \myref{zsoln}.
I.e., fixing one of these defines a $7+1$ dimensional hypersurface
which defines the worldvolume of the D7 brane.  This is then
substituted into the Dirac-Born-Infeld (DBI) action and expanded around fluctuations
in this embedding to define the ``mesons'' of the theory.

Our interest in this topic began with \cite{Gherghetta:2006yq,Gabella:2007cp}.  In those
papers we were looking for a string theoretic basis for warped extra dimension models
of the Randall-Sundrum type \cite{Randall:1999ee}.
One of the issues in these models is stabilizing the hierarchy.  We believed that
there is a string theoretic version of this, based on the work of Giddings, Kachru
and Polchinski \cite{Giddings:2001yu}. By contrast, effective field theoretic
approaches to the problem of stabilizing the extra dimension were
initiated by Goldberger and Wise in \cite{Goldberger:1999uk}.

The D7 is embedded into the spacetime by the equation
\beq
z_1 = r^{3/2} \sin \frac{\theta_1}{2} \sin \frac{\theta_2}{2} 
e^{ \frac{i}{2} ( \psi - \phi_1 - \phi_2 ) }
= \mu > 0
\eeq
Here, $\mu$ is an adjustable (real) parameter of the embedding
which is related to how far towards the tip of the conifold the
D7 branes extend.  This leads to $r$ and $\psi$ being functions
of the other coordinates:
\beq
r = r_0(\theta_i), \quad
\psi = \psi_0(\phi_i)
\eeq
In particular,
\beq
r_0^{3/2} \sin \frac{\theta_1}{2} \sin \frac{\theta_2}{2} = \mu,
\quad
\psi_0 = \phi_1 + \phi_2
\eeq
These equations are important in the analysis of fluctuations
of the D7 branes, which lead to modes of excitation in the effective
theory.  The fluctuations of the embedding are parameterized by
\beq
r = r_0 ( 1 + \chi ), \quad \psi = \psi_0 + 3 \eta
\eeq
Here, $\chi$ and $\eta$ represent the fluctuations and are functions
of the D7 worldvolume coordinates $\xi^a$, $a=0,1,\ldots,7$:
\beq
\chi = \chi(\xi^a), \quad
\eta = \eta(\xi^a)
\eeq
We will identify $\xi^a$ with the 8 coordinates $x^\mu$ (4d spacetime), $\theta_i$
and $\phi_i$.

The embedding metric can be obtained in a couple of ways.  One is through the formula
\beq
G_{ab} = g_{MN} \frac{\p X^M}{\p \xi^a} \frac{\p X^N}{\p \xi^b}
\eeq
where $X^M=X^M(\xi)$ are the 10d coordinates describing the embedding of the probe D7 brane.
An alternative is to start with the 10d metric \myref{10dmetric} and
substitute in the expressions
\beq
d r &=& \frac{\p r_0}{\p \theta_i} d\theta_i
+ r_0 ( \frac{\p \chi}{\p x^\mu} dx^\mu + \frac{\p \chi}{\p \theta_i} d\theta_i
+ \frac{\p \chi}{\p \phi_i} d\phi_i )
\nnn
d \psi &=& d\phi_1 + d\phi_2 + 3 ( \frac{\p \eta}{\p x^\mu} dx^\mu + \frac{\p \eta}{\p \theta_i} d\theta_i
+ \frac{\p \eta}{\p \phi_i} d\phi_i )
\label{drdpsi}
\eeq
This can then be substituted into the 10d metric \myref{10dmetric} above to obtain
an equation for the 8d metric of the D7 brane embedding.

Working in the near-horizon limit of \myref{10dmetric} $r \to 0$, and substituting
the expressions \myref{drdpsi}, we obtain the 8d metric including fluctuations:
\beq
ds_8^2 = \frac{r_0^2}{L^2} (1 + \chi)^2 \eta_{\mu\nu} dx^\mu dx^\nu
+ \frac{L^2}{r_0^2} (1 + \chi)^{-2} dr^2 + d{\tilde s}^2_{T^{1,1}} 
\eeq
Here, $d{\tilde s}^2_{T^{1,1}}$ takes into account the expression for $d\psi$ in
\myref{drdpsi} above, and $dr$ should also use the corresponding
expression from \myref{drdpsi}.
The nonzero elements of the 8d metric are then found to be:
\beq
G_{\mu\nu} &=& \frac{r_0^2}{L^2} (1 + \chi)^2 \eta_{\mu\nu} \nnn
G_{\theta_i \theta_i} &=&  \frac{L^2}{r_0^2} (1 + \chi)^{-2} \( \frac{\p r_0}{\p \theta_i} + \frac{\p \chi}{\p \theta_i} \)^2
+ \frac{1}{6} +  \( \frac{\p \eta}{\p \theta_i} \)^2 \nnn
G_{\theta_1 \theta_2} &=& G_{\theta_2 \theta_1} =
\frac{L^2}{r_0^2} (1 + \chi)^{-2} \( \frac{\p r_0}{\p \theta_1} + \frac{\p \chi}{\p \theta_1} \) 
\( \frac{\p r_0}{\p \theta_2} + \frac{\p \chi}{\p \theta_2} \)
+ \frac{\p \eta}{\p \theta_1} \frac{\p \eta}{\p \theta_2} 
\eeq
This is then substituted into the DBI effective action:
\beq
S_\text{DBI} = -\tau_7 \int d^4 x ~ d^2 \theta ~ d^2 \phi ~
\sqrt{ \det [ G + \varphi^\star(B) + 2 \pi \alpha' F ] }
\eeq
where $\varphi^\star(B)$ is the pullback of the antisymmetric tensor (which
will not be important for our purposes), $F$ is the $U(1)$ field strength
associated with the charge of the D7 brane and $\tau_7 = (2 \pi)^{-7} \alpha'^4 g_s^{-1}$
is the brane tension.

By a sequence of redefinitions and approximations, angular coordinates
can be integrated out and one obtains for an effective, quadratic action:
\beq
S(\chi) \approx -2 \pi^2 L^{-5} \tau_7 \int d^4 x \int_R^\infty dr
\bigg\{ \frac{r}{L} \eta^{\mu\nu} \p_\mu \chi \p_\nu \chi
+ \frac{r^5}{L^5} ( \p_r \chi )^2
- \frac{15}{4 L^2} \frac{r^3}{L^3} \chi^2 \bigg\}
\eeq
Here, $L$ is the $AdS_5$ radius.
A similar equation holds for $\eta$.
This is the action of a conformally coupled scalar in $AdS_5$.
Thus we see that an effective five-dimensional theory can be
derived from D7 probe branes embedded into the KW geometry
and that it has a standard form.

\section{Klebanov-Witten duality}
Klebanov and Witten showed the duality of the conifold compactified
string theory to a superconformal gauge theory \cite{Klebanov:1998hh}.
Here we review some key aspects of that duality.

For the conifold, we can describe it in terms of a quaternion:
\beq
Z = \frac{1}{\sqrt{2}} \sum_{n=1}^4 w_n \s^n
\eeq
Here, $\s^4$ is the unit matrix multiplied by $i = \sqrt{-1}$.
Then the conifold equation can be described by:
\beq
\det Z = 0
\label{detZzero}
\eeq
This can be related to the dual gauge theory by writing $Z$ as:
\beq
Z = \begin{pmatrix} A_1 B_1 & A_1 B_2 \cr
A_2 B_1 & A_2 B_2 \end{pmatrix}
\eeq
Then the superpotential is given by
\beq
W = \lambda \det Z
\eeq
which is a function of the chiral superfields $A_i$ and $B_i$.
Here, $\lambda$ is a parameter.  In supergravity, one of the conditions\footnote{This
particular condition is related to the supersymmetric variation of the gravitino field.}
for a supersymmetric ground state is $W = 0$.  Hence we recover
the conifold equation from this condition.  In actuality, if $A_i$ and $B_i$
are all commuting variables, then it is identically true that
\beq
\det Z = A_1 B_1 A_2 B_2 - A_1 B_2 A_2 B_1 = 0
\eeq
Thus the parameterization of $Z$ in terms of these variables
automatically generates the conifold condition $\det Z = 0$.
However, if $A_i$ and $B_i$ are matrices, as in the $SU(N) \otimes SU(N)$
gauge theory (where they are bi-fundamentals), then
\beq
W = \lambda \tr ( A_1 B_1 A_2 B_2 - A_1 B_2 A_2 B_1 )
\eeq
does not vanish identically.  In this case we have a nontrivial
superpotential and the vacuum conditions correspond to the conifold.
In particular, if on the moduli space $A_i$ and $B_i$ are all
diagonal matrices, then the condition $W=0$ is satisfied.

The parameterization in terms of $Z$ also reveals symmetry.  We note
that \myref{detZzero} allows for the transformation
\beq
Z \to U Z V^\dagger
\eeq
where $U$ and $V$ are $SU(2)$ group matrices.  This is
because
\beq
\det U = \det V = 1
\eeq
So, there is an $SU(2) \otimes SU(2)$ isometry for the conifold.
Also, one can rephase $Z$:
\beq
Z \to e^{i \varphi} Z
\eeq
where $\varphi$ is a real parameter.  Thus there is an additional
$U(1)$ isometry.  Of course one can also see these
isometries from the equation \myref{wsalladdtozero}.
One simply notes that $SO(4) \simeq SU(2) \otimes SU(2)$.
Obviously equation \myref{wsalladdtozero} is invariant under
a four-dimensional rotation.  Similarly, the $z$s can
all be simultaneously rephased by $e^{i \varphi}$.

\section{Klebanov-Strassler}
The Klebanov-Strassler (KS) construction \cite{Klebanov:2000hb} has been reviewed
previously in Strassler \cite{Strassler:2005qs}
and Gwyn and Knauf \cite{Gwyn:2007qf}.  For details beyond
what is found in this brief section, see those references.

Recall that in the Klebanov-Witten construction the dual gauge theory
has gauge group $G = SU(N) \otimes SU(N)$.  By contrast, KS
has a gauge group $G = SU(N+M) \otimes SU(N)$ which undergoes a cascade
of Seiberg duality transformations \cite{Seiberg:1994pq} as one descends in energy.  Because
of this, the theory confines in the IR.  The consequent chiral symmetry
breaking resolves the IR singularity of the KW construction. 

On the brane side, Klebanov-Strassler corresponds to $N$ D3 branes
and $M$ wrapped D5 branes.  This takes advantage of the fact that
$T^{1,1}$ is topologically equivalent to $S^3 \times S^2$.  Thus
there is an $S^2$ to wrap, leaving effectively branes with three
spatial dimensions (fractional D3 branes).  Often it is said that
the D5 branes wrap a two-cycle.

\subsection{Geometry}
In terms of the geometry, the KS case has a simple
modification of the conifold equation:  the sum is
nonzero.  That is,
\beq
w_1^2 + w_2^2 + w_3^2 + w_4^2 = \e^2
\label{deformed}
\eeq
with $\e$ some nonzero number.  This means that there
is a minimal radius allowed.  Thus the tip of the conifold
is rounded off and is no longer singular.
This space is referred to as the deformed conifold.
Because warp factors will ultimately be applied to this
geometry, KS is built on the warped deformed conifold.

One of the nice features
about the KS construction is that it
makes explicit what was suspected intuitively all along:
the radial coordinate on the gravity side maps to the RG
scale on the gauge theory side.  This occurs through a duality
cascade and is a result of the fractional D3 branes.

It is interesting that the two principle modifications of the conifold can be
related to each other.  In the deformed conifold the size of the 3-cycle
is described by the complex structure modulus.  On the other hand in the
resolved conifold the size of the 2-cycle corresponds to the K\"ahler
modulus.  These are mapped to each other in the topological string
context \cite{Gopakumar:1998ki}.  In that case the large $N$ Chern-Simons
theory on $S^3$ is shown to be dual to the topological string
on the $S^2$ blow-up of the conifold.
This suggests a geometric transition may also be possible
between the deformed and resolved geometries in the full
string theory.  It would be interesting to see what this does to
the dual gauge theory.

The resolved conifold is described as follows.  Translating
the notation in Elituv \cite{Elituv:2018byg} to the one used here,
\beq
\begin{pmatrix} z_1 & -z_3 \cr
-i z_4 & i z_2 \end{pmatrix}
\begin{pmatrix} \eta_1 \cr \eta_2 \end{pmatrix}
\eeq
for some real parameters $\eta_1$ and $\eta_2$.
It can be seen that this is quite different from \myref{deformed},
so it is a little surprising that there may be a duality here.

KS has the additional features beyond those of KW:
fractional D3 branes, duality cascades, IR resolution of the singularity
and chiral symmetry breaking.  These are all captured
in the modified geometry, which we now describe.

The KS metric is conveniently described in terms of 1-forms:
\beq
ds_{10}^2 &=& A^2(\tau) \eta_{\mu\nu} dx^\mu dx^\nu
+ B^2(\tau) d\tau^2 + C^2(\tau) (g^5)^2
\ddd + D^2(\tau) [ (g^3)^2 + (g^4)^2 ]
+ E^2(\tau) [ (g^1)^2 + (g^2)^2 ]
\eeq
The one-forms in the above expressions can be related to the original KW
coordinates:
\beq
g^1 &=& \frac{e^1 - e^3}{\sqtw}, \quad g^2 = \frac{e^2 - e^4}{\sqtw} \nnn
g^3 &=& \frac{e^1 + e^3}{\sqtw}, \quad g^4 = \frac{e^2 + e^4}{\sqtw} \nnn
g^5 &=& e^5
\eeq
where
\beq
e^1 &=& -\sin\theta_1 d\phi_1, \quad e^2 = d\theta_1 \nnn
e^3 &=& \cos\psi \sin\theta_2 d\phi_2 - \sin\psi d\theta_2 \nnn
e^4 &=& \sin\psi \sin\theta_2 d\phi_2 + \cos\psi d\theta_2 \nnn
e^5 &=& d\psi + \cos\theta_1 d\phi_1 + \cos\theta_2 d\phi_2
\eeq

Based on various considerations about the allowed fluxes in the theory,
this is further specialized to \cite{Klebanov:2000hb}:
\beq
ds_{10}^2 &=& h^{-1/2}(\tau) \eta_{\mu\nu} dx^\mu dx^\nu
+ h^{1/2}(\tau) ds_6^2 
\label{KSgen}
\\
ds_6^2 &=& \half \e^{4/3} K(\tau) \bigg[ \frac{1}{3 K^3(\tau)}
(d\tau^2 + (g^5)^2) + \cosh^2(\tau/2) [ (g^3)^2 + (g^4)^2 ]
\ddd \quad + \sinh^2(\tau/2) [ (g^1)^2 + (g^2)^2 ] \bigg] 
\label{KS6d}
\\
K(\tau) &=& \frac{(\sinh(2\tau)-2\tau)^{1/3}}{2^{1/3} \sinh\tau}
\eeq 
The warp factor $h(\tau)$ is obtained from solving the supergravity
equations of motion, yielding:
\beq
h(\tau) = \alpha \frac{2^{2/3}}{4} \int_\tau^\infty dx ~
\frac{x \cosh x - 1}{\sinh^2 x}
(\sinh(2x) - 2x)^{1/3}
\label{KSH}
\eeq
Obviously this is far more intricate than KW, reflecting that this is a supergravity
dual of a theory with confinement and spontaneous chiral symmetry breaking (gaugino
condensation).  It is
therefore more realistic as a starting point for phenomenology.

It is interesting to compare KS to KW in detail.  For instance, \myref{KSgen}
versus \myref{10dmetric} looks quite similar.  But the 6d part in
\myref{KS6d} is quite a bit more complicated.  Also, the function $H(r)$
in \myref{KWH} is quite simple, just reflecting a stack of D3 branes
at the origin $r=0$, whereas the corresponding function \myref{KSH} in KS
is rather involved, reflecting wrapped D5 branes at all different
values of the radial coordinate $\tau$.  In the dual gauge theory
this latter complicated $\tau$ dependence is exhibited by the
Seiberg duality cascade.

In KS, we find the relation $r^3 \sim \e^2 e^\tau$.  Based on this, we could
choose to embed a probe D7 brane in a way
that is analogous to $z_1=\mu$ in KW, namely:
\beq
\e e^{\tau/2} e^{\frac{i}{2} ( \psi - \phi_1 - \phi_2 ) }
\sin \frac{\theta_1}{2} \sin \frac{\theta_2}{2} = \mu
\eeq
There is a question however:  is it a supersymmetric embedding?  To
answer this question requires a detailed investigation that we
will not address in this review.

\subsection{Forms and couplings}
The background field strength $F_3$ is governed by the number of
D5 branes $M$ according to \cite{Klebanov:2000hb}:
\beq
F_3 &=& M \omega_3 \nnn
\omega_3 &=& \half d \psi \wedge ( \sin \theta_1 d \theta_1 \wedge d \phi_1
- \sin \theta_2 d \theta_2 \wedge d \phi_2 ) \nnn
&& + \half d \phi_1 \wedge d \phi_2 \wedge ( \cos \theta_1 \sin \theta_2 d \theta_2
+ \sin \theta_1 \cos \theta_2 d \theta_1 )
\eeq
The gauge couplings in the two factors of the gauge group are
related by
\beq
\frac{1}{g_1^2} - \frac{1}{g_2^2} \sim
\frac{1}{g_s} \( \int B_2 - \half \)
\label{g1g2eq}
\eeq
where $g_s$ is the string coupling and $B_2$ is the NS-NS 2-form.
The latter is given by
\beq
B_2 &=& 3 g_s M \omega_2 \ln \( \frac{r}{r_0} \) \nnn
\omega_2 &=& \half ( \sin \theta_1 d \theta_1 \wedge d \phi_1
- \sin \theta_2 d \theta_2 \wedge d \phi_2 )
\label{B2eq}
\eeq
$r_0$ is a reference scale.
Thus we see that the difference in gauge couplings depends on the
radial coordinate $r$, corresponding under the duality to a
renormalization group scale.  This is one of the interesting
things about holography:  in a theory with a running coupling,
the extra dimension corresponds to the renormalization group
scale.  Here it is important to have an example like KS, where
the gauge theory is not conformal.

In the KW and KS constructions there are two gauge couplings, $g_1$ and $g_2$,
corresponding to the two factors of the gauge group.  In KW, they can take
any values without destroying the conformal invariance.  In KS we
have to keep in mind the duality cascade.  What do these couplings
correspond in the dual SUGRA?  Klebanov and Strassler discuss this and
we mention a few details here because the story is interesting.

For instance in the $M=0$ scenario (no fractional D3 branes, i.e., just KW), 
we have the formulae:
\beq
\frac{1}{g_1^2} + \frac{1}{g_2^2} \sim e^{-\phi}
\eeq
and
\beq
\frac{1}{g_1^2} - \frac{1}{g_2^2} \sim e^{-\phi} \[ \int_{S^2} B_2 - \frac{1}{2} \]
\eeq
equivalent to \myref{g1g2eq} above.
In the T-dual Type IIA picture, it is related to the positions of NS5 branes \cite{Klebanov:2000hb}:
\beq
\frac{1}{g_1^2} = \frac{l_6-a}{g_s}, \quad \frac{1}{g_2^2} = \frac{a}{g_s}
\eeq
Here, $l_6$ is the size of the compact dimension upon which things are T-dualized,
and $a$ is the separation of the two NS5 branes.

In the KW case, $\int_{S^2} B_2$ is normalized to that its period is 1.  By contrast
in KS with $M \not= 0$ we have
\beq
\int_{S^2} B_2 \sim M e^\phi \ln(r/r_0)
\eeq
as can be seen from \myref{B2eq} above,
so that a logarithmic running occurs for the gauge couplings as we
follow the duality cascade down to small $r$.

\section{Other generalizations of the conifold}
In \cite{Elituv:2018byg}, Elituv explores both K\"ahler and non-K\"ahler 
3-complex-dimensional manifolds (locally $\Cbf^3$)
that are generalizations of the various constructions such as Klebanov-Witten, Klebanov-Strassler, etc.
He begins with the forms $dr, d\psi, d\theta_i, d\phi_i$ and imposes that it is a complex manifold by
computing $d\Omega=0$ where $\Omega$ is a rather general $(3,0)$ form.  Then, the K\"ahler
condition is tackled using $d(e^{2\phi} J)$ where $J$ is a rather general $(1,1)$ form
and $\phi$ is the dilaton.  From this he derives conditions on the various coefficient functions,
including some differential equations that they should satify.  This opens up the possibilty
of many new manifolds that could serve as the basis for holographic duals to some $\Ncal=1$
supersymmetric gauge theory.

\subsection{Conditions for supersymmetry}
To establish that a conifold-like theory has $\Ncal=1$ supersymmetry,
Elituv explains that we need to check three things:
\ben
\item
The complex manifold has a (3,0) form $\Omega$.  We need to then require
that the manifold be closed, $d \Omega = 0$.  This imposes constraints
on the components $\Omega_{abc}$.  Solving these constraints then
tells us something about the (3,0) form.  But the manifold must be
consistent with the existence of a solution.  So it also tells us
something about the manifold.
\item
We construct the 3-form $G_3 = F_3 - i e^{i \phi} H_3$.
Here, $F_3$ is the field strength associated with a 2-form gauge potential
and $H_3$ is given by $H_3 = d B_2$ where $B_2$ is the usual 2-form
associated with string theory.  $\phi$ is the dilaton.
Next we demand that the the 3-form $G_3$ is a (2,1) form.
That is, all of the non-(2,1) form components need to vanish.
This requires that we specify the complex basis of the manifold,
and again tells things about both the form and the manifold.
\item
Lastly, we have require that the (2,1) form is primitive.
This means that $J \wedge G_3 = 0$, where
\beq
J = i g_{a {\bar a}} dz^a \wedge d {\bar z}^{{\bar a}}
\eeq
is the K\"ahler form.
\een

\subsection{A more general metric}
As elucidated in Elituv, the more general metric in the case of
deformed and resolved conifolds can be described by:
\beq
ds^2 &=& {\cal G}_1 dr^2 + {\cal G}_2 ( d\psi + \cos \theta_1 d\phi_1
+ \cos \theta_2 d\phi_2 )^2
\ddd + {\cal G}_3 ( d\theta_1^2 + \sin^2 \theta_1 d\phi_1^2 ) 
+ {\cal G}_4 ( d\theta_2^2 + \sin^2 \theta_2 d\phi_2^2 )
\ddd + {\cal G}_5 \cos \psi ( d \theta_1 d \theta_2
- \sin \theta_1 \sin \theta_2 d \phi_1 d \phi_2 )
\ddd + {\cal G}_6 \sin \psi ( \sin \theta_1 d \phi_1 d\theta_2
+ \sin \theta_2 d\phi_2 d \theta_1 )
\label{genmetric}
\eeq
The functions ${\cal G}_1, \ldots, {\cal G}_6$ then need to be
specified in either case.  Furthermore, this form allows for
various generalizations.  Elituv calls this a ``non-K\"ahler
resolved deformed conifold.''

Elituv reports that the deformed conifold corresponds to
\beq
{\cal G}_1 &=& \frac{\gamma + (r^2 \gamma' - \gamma) \( 1 - \frac{\mu^4}{r^4} \) }{
r^2 \( 1 - \frac{\mu^4}{r^4} \)} \nnn
{\cal G}_2 &=& \frac{1}{4} \[ \gamma - (r^2 \gamma' - \gamma) \( 1 - \frac{\mu^4}{r^4} \) \]
\nnn
{\cal G}_3 &=& {\cal G}_4 = \frac{\gamma}{4}, \quad
{\cal G}_6 = \frac{\mu^2 \gamma}{2 r^2}
\eeq
Here $\mu^2$ is a deformation parameter related to $\e^2$ above.
$\gamma$ is some function of the radial coordinate $r$.
$\gamma' = d \gamma / d(r^2)$.

Note that for D5 branes, the natural gauge potential that couples
to them is a 6-form $C_6$.  The corresponding field strength is
$F_7 = d C_6$.  The Hodge dual of this is $F_3 = {}^* F_7$.
This is the field above that appears in $G_3$.

In the next subsection, we describe how Elituv is able to
construct the metric, the (1,1) form $J$ and the (3,0) form $\Omega$
from a basis of complex forms ${\cal E}_1$, ${\cal E}_2$ and ${\cal E}_3$.
This involves both the functions ${\cal G}_1, \ldots, {\cal G}_4$
and some free parameters $\alpha_1, \ldots, \alpha_4$ and
$\beta_1, \ldots, \beta_4$.  He can then study the conditions on
these quantities if the manifold is K\"ahler, $d(e^{2\phi} J)=0$, and/or complex,
$d\Omega=0$.  Having started with a rather general set-up, he can
construct manifolds that are more general than the ones appearing
in the original KW and KS formulations.  It is an interesting question
how these modifications would be reflected in the dual gauge theory.

\subsection{Complex geometry}
\label{complex}
The conifold has the nice feature that it illustrates a system describable by
complex geometry.  This is already apparent from the fact that we utilize
four complex coordinates $z_i$ satisfying \myref{zeq}.  This is similar to
how we often describe a sphere in terms of embedding coordinates in a higher
dimensional space, subject to a constraint equation.  Here we begin with
four complex dimensions, equivalent to eight real dimensions.  But we impose
one constraint equation, which happens to be complex.  This reduces us to
a three-complex-dimensional subspace, equivalent to a six-real-dimensional subspace.
As in the case of a sphere, we begin with a flat metric in the higher
dimensional space:
\beq
ds^2 = \sum_{i=1}^4 d \zbar_i dz_i
\eeq
To get to the metric written in terms of the Gaussian coordinates $r,\psi,\theta_{1,2},\phi_{1,2}$
we substitute in the solutions \myref{zsoln} to the constraint equation \myref{zeq}.

In the theory of the conifold, one deals with the
left-invariant Maurer-Cartan forms:
\beq
\s_1 &=& \cos \psi_1 d\theta_1 + \sin\psi_1 \sin\theta_1 d\phi_1 \nnn
\s_2 &=& -\sin\psi_1 d\theta_1 + \cos\psi_1 \sin\theta_1 d\phi_1 \nnn
\s_3 &=& d\psi_1 + \cos\theta_1 d\phi_1 \nnn
\Sigma_1 &=& \cos \psi_2 d\theta_2 + \sin\psi_2 \sin\theta_2 d\phi_2 \nnn
\Sigma_2 &=& -\sin\psi_2 d\theta_2 + \cos\psi_2 \sin\theta_2 d\phi_2 \nnn
\Sigma_3 &=& d\psi_2 + \cos\theta_2 d\phi_2 \nnn
\eeq
where $\psi_1 = \psi_2 = \half \psi$.
Using these, one can build up the various structures that arise.
These are discussed for instance in Section 3.1 of \cite{Elituv:2018byg}.

Note that $\s_1$ and $\s_2$ can be obtained as 2d rotations of the more
basic forms $d\theta_1$ and $\sin \theta_1 d\phi_1$ that appear in the
metric of $S^2$:
\beq
\begin{pmatrix} \s_1 \cr \s_2 \end{pmatrix}
= \begin{pmatrix} \cos \psi_1 & \sin \psi_1 \cr
-\sin \psi_1 & \cos \psi_1 \end{pmatrix}
\begin{pmatrix} d\theta_1 \cr \sin \theta_1 d \phi_1 \end{pmatrix}
\eeq
A similar relation holds for $\Sigma_1$ and $\Sigma_2$. 

Elituv relates these left-invariant Maurer-Cartan forms
back to his rather general metric \myref{genmetric} after
various manipulations and assumptions.  Thus these forms
can be viewed as building blocks for constructing modifications
of the original conifold geometry.

Elituv uses these forms to compose other forms that are
particular useful for describing his geometry.  For instance,
he has:
\beq
e_1 &=& \sqrt{ {\cal G}_1 } dr
\nnn
e_2 &=& \sqrt{ {\cal G}_2 } ( \s_3 + \Sigma_3 )
\nnn
e_3 &=& \sqrt{ {\cal G}_3 } ( \alpha_1 \s_1 + \beta_3 \Sigma_1 )
\nnn
e_4 &=& \sqrt{ {\cal G}_4 } ( \alpha_2 \s_2 - \beta_4 \Sigma_2 )
\nnn
e_5 &=& \sqrt{ {\cal G}_3 } ( \beta_1 \s_1 + \alpha_3 \Sigma_1 )
\nnn
e_6 &=& \sqrt{ {\cal G}_4 } ( -\beta_2 \s_2 + \alpha_4 \Sigma_2 )
\eeq

From these he then constructs complex forms:
\beq
{\cal E}_1 = e^{-\phi} ( e_2 + i e_1 ), \quad
{\cal E}_2 = e^{-\phi} ( e_3 + i e_4 ), \quad
{\cal E}_3 = e^{-\phi} ( e_5 + i e_6 )
\eeq
These can then be used to form a 6d metric.
\beq
e^{-2 \phi} ds_6^2 = {\cal E}_1 \otimes {\bar {\cal E}}_1
+ {\cal E}_2 \otimes {\bar {\cal E}}_2
+ {\cal E}_3 \otimes {\bar {\cal E}}_3
\eeq
Elituv also uses these in the construction of the (1,1) form:
\beq
J = - \frac{i}{2} ( {\cal E}_1 \wedge {\bar {\cal E}}_1
+ {\cal E}_2 \wedge {\bar {\cal E}}_2
+ {\cal E}_3 \wedge {\bar {\cal E}}_3 )
\eeq
Likewise, he obtains the (3,0) form from these:
\beq
\Omega = e^{3 \phi} {\cal E}_1 \wedge {\cal E}_2 \wedge {\cal E}_3
\eeq
Thus we have a general prescription for how to build
up the relevant structures in the complex geometry, based
on the left-invariant Maurer-Cartan forms.

The conifold is a noncompact Calabi-Yau manifold.
This means it has $SU(3)$ holonomy since it has 3 complex
dimensions.  The result is
$\Ncal=1$ supersymmetry in the 4d low energy theory
since the manifold has a single Killing spinor.
Now one might ask:  if it is noncompact, how does one get 4d physics?  One might
think that one has to
completly compactify the 6 extra dimensions
of superstring theory in order to get 4d physics.
So, how does it work?  Part of the answer is that we have D3 branes and wrapped D5 branes (on the
$\sim S^2$ part of $T^{1,1}$).  So of course the worldvolume theory on these branes is 4d.
But it is more than that, as we can see from the effective 5d picture derived
in \cite{Gherghetta:2006yq}.  There we see that we have effectively a Randall-Sundrum type setup with
a UV brane and an IR brane.  It is well-known that this will produce an effective
4d theory at low energy.  Evidence of the extra 5th dimension begins to appear at the TeV
scale through KK excitations.  Thus there is a compactification of 5 dimensions on
$T^{1,1}$, and the remaining extra dimension is dealt with in a Randall-Sundrum
sort of way.

\section{Lattice theory}
The most straightforward lattice fermion is the Wilson fermion, which has an action
\beq
S = \sum_x \{ \psib \gamma_\mu D_\mu \psi + \half r a \psib D^2 \psi + m \psib \psi \}
\eeq
The role of the Wilson ``mass term'' $\psib D^2 \psi$ is to lift spectral doublers. 
$a$ is the lattice spacing and $r$ is a dimensionless constant.  The
factor of $\half$ is conventional.  Even with $m=0$ chiral
symmetry is violated explicitly, due to the $\psib D^2 \psi$ term.  
This causes problems for preservation of global
symmetries in supersymmetric theories, as we now detail.

In two-component notation,
the part of the SUSY action for fermions coming from the superpotential
is given by\footnote{Often, the index on the fermion and other fields
would be raised, to make the K\"ahler geometric interpretation more
clear.  However, we will not use this notation here in our relatively
brief discussion.}
\beq
-\half W_{ij} \chi_i \chi_j -\half \overline{W}_{\ibar \jbar} \chib_\ibar \chib_\jbar
\eeq
Here, $W_{ij} = \p^2 W/\p \phi_i \p \phi_j$ is the second derivative of the superpotential.
The example that we will refer to is the one for the KW and KS theories:
\beq
W = \lambda \e_{ij} \e_{kl} \tr ( A_i B_k A_j B_l )
\label{KSspot}
\eeq
Thus for instance the SUSY Lagrangian will contain a term
\beq
 -\lambda \e_{ij} \e_{kl} \tr ( \chi_{Ai}^\alpha B_k \chi_{Aj \alpha} B_l )
\eeq
where $\alpha = 1,2$ is the spinor index on the two-component fermions,
$\chi_{Ai}$ being the fermionic partner of the scalar $A_i$.

In order to understand the chiral symmetry, and how the Wilson ``mass term'' 
will impact it, it is useful to go over to a four-component notation.
For this purpose the fermions are grouped into Majorana fermions
\beq
\Psi_{MAi} = \binom{\chi_{Ai}}{\chib_{A \ibar}}, \quad
\Psi_{MBi} = \binom{\chi_{Bi}}{\chib_{B \ibar}}
\eeq
The superpotential \myref{KSspot} has a global $SU(2)_A \times SU(2)_B \times U(1)$ symmetry.
The two SU(2)s are obvious, and the U(1) acts on the superfields according
to $A_i \to e^{i\alpha} A_i$, $B_i \to e^{-i \alpha} B_i$. 

Focusing on the $\s^3$ part of the first SU(2), we have
\beq
\chi_{A1} \to e^{i\alpha} \chi_{A1}, \quad
\chi_{A2} \to e^{-i\alpha} \chi_{A2}, \quad
\chib_{A1} \to e^{-i\alpha} \chib_{A1}, \quad
\chib_{A2} \to e^{i\alpha} \chib_{A2}
\eeq
In terms of the Majorana fermions, this is
\beq
\Psi_{MA1} \to e^{-i \alpha \gamma_5} \Psi_{MA1}, \quad
\Psi_{MA2} \to e^{i \alpha \gamma_5} \Psi_{MA2}
\eeq
The Wilson ``mass term''
\beq
\Psib_{MA1} D^2 \Psi_{MA1} + \Psib_{MA2} D^2 \Psi_{MA2}
= \Psi_{MA1}^T C D^2 \Psi_{MA1} + \Psi_{MA2}^T C D^2 \Psi_{MA2} 
\eeq
badly violates this symmetry.  However, from the last expression,
it can be seen that an SO(2) subgroup of the SU(2) is preserved:
\beq
\binom{\Psi_{MA1}}{\Psi_{MA2}} \to 
\begin{pmatrix} \cos \theta & \sin \theta \cr
-\sin \theta & \cos \theta \end{pmatrix}
\binom{\Psi_{MA1}}{\Psi_{MA2}} 
\eeq
Importantly, this symmetry is restrictive on the necessary counterterms.

Thus we see that trying to formulate either the KW or the KS theories
on the lattice using the Wilson discretization for fermions will
violate the $SU(2) \times SU(2) \times U(1)$ flavor symmetry
of these models.  Because it is broken by the regulator, we do
not expect it to be restored in the IR without tuning counterterms.
The number of such counterterms will be large and they would have
to be fine-tuned nonperturbatively, which would require many simulations.
This is an example of the types of challenges that are faced when
trying to study these theories from first principles using
numerical methods.

\section{The D1-D5 system}
Although the phenomenological implications are unclear,\footnote{Since the dual
CFT is two-dimensional, it may be that the D1-D5 system and its close
relatives could have applications in condensed matter systems close to
a second order critical point.  This could include static properties
of 2+1 dimensional systems or dynamical properties of 1+1 dimensional systems.}
 the D1-D5 system on $T^4$ is of
particular interest in the field of holography, such as the recent work \cite{Costello:2020jbh}.  It is supposed to be dual to 
the supersymmetric gauge theory on $AdS_3 \times S^3 \times T^4$.  Indeed, we have
previously studied an approach that may realize a latticization of this system \cite{Giedt:2011zza}.
More generally, this setup is just one example of a whole class of
2d conformal theories that are dual to 6d supergravity formulated on
$AdS_3 \times S^3$.  Early stages of such an analysis have been
started by de Boer \cite{deBoer:1998kjm}, but much more work remains to be done.  One
can ask questions such as:  have we learned anything from the bootstrap
studies that might shed light on these types of dualities?
How much is really known about these 6d supergravities?
Are there more elegant formulations of them, analogous to
the K\"ahler U(1) superspace for N=1 sugra in 4d?  (See \cite{Binetruy:2000zx}
and references therein.)

Six-dimensional Calabi-Yau manifolds $K$ are an important compactification
target within string theoretic studies, because of the nice
properties of these spaces.  For this reason there is also
some attention on compactification of F-theory (which is 12d)
on $AdS_3 \times S^3 \times K$ and M-theory (which is 11d)
on $AdS_3 \times S^2 \times K$.  The common element is $AdS_3$,
which means that these theories will have dual 2d CFTs.  Thus
they generally fall into the same class as the D1-D5 based
theories.

In \cite{Costello:2020jbh}, a connection is made to twisted supersymmetry.
This is especially challenging on the gravity side of the duality, since it requires
supergravity to be twisted.  Nevertheless, this is an interesting direction
in light of how many of the supersymmetric lattices are formulated in
a twisted framework.

In \cite{Eloy:2021fhc} supersymmetry is broken
in systems analogous to D1-D5.  This is accomplished through
supergravity in geometries that are topologically
$AdS_3 \times S^3$.  In this particular situation, the
$S^3$ is replaced by a squashed sphere.
These surfaces preserve an $SO(2) \otimes SO(2)$ isometry
group of the original $SO(4)$ isometry group of $S^3$.
These calculations rely heavily on exceptional field theory.

For the D1-D5 system with D5s wrapped on the four-dimensional
manifold K3, the metric is given by \cite{deBoer:1998kjm}
\beq
\frac{ds^2}{\alpha'} &=& \frac{U^2}{\ell^2}
( -dt^2 + (dx^5)^2 ) 
+ \frac{\ell^2}{U^2} dU^2
\ddd
\quad
+ \ell^2 d\Omega_3^2 
+ \sqrt{\frac{Q_1}{vQ_5}} ds_{K3}^2
\eeq
Here, $Q_1$ is the number of D1 branes and $Q_5$ is the number of D5 branes.
The $d\Omega_3^2$ is associated with an $S^3$.  Note that $U$ is a sort of
radial coordinate.  The parameter $\ell$ is a length scale.  It is
given by $\ell^2 = g_6 \sqrt{N}$, where $N=Q_1 Q_5$ and $g_6$ is the
six-dimensional gauge coupling, associated with reducing the ten-dimensional
theory using compactification on K3. 

$AdS_3$ can be expressed through the metric
\beq
ds^2 = \ell^2 ( - \cosh^2 \rho d\tau^2
+ \sinh^2 \rho d\phi^2 + d\rho^2 )
\eeq
Here, $\ell$ is the AdS radius (inverse of AdS curvature).  The masses
of primary fields on $AdS_3$ are given by \cite{deBoer:1998kjm}:
\beq
m^2 \ell^2 = ( h + {\bar h} ) ( h + {\bar h} - 2 )
\eeq
where $h$ and ${\bar h}$ are the conformal weights at the boundary.
This is typical of the AdS/CFT correspondence.
We see that there is a relationship between the mass of the
field in anti-de-Sitter space and the conformal weights in
the CFT at the boundary.  A similar relation holds in
more phenomenological applications that relate $AdS_5$ to 4d effective
field theory in terms of Kaluza-Klein decomposition.

Associated with this geometry are Virasoro operators.  Those corresponding
to the $SL(2,\Rbf)$ subalgebra of the 2d conformal group are given by:
\beq
L_0 &=& i \p_u \nnn
L_{-1} &=& i e^{-i u} \( \coth ( 2 \rho ) \p_u
- \frac{1}{ \sinh (2 \rho ) } \p_v
+ \frac{i}{2} \p_\rho
- \frac{i}{2} s \coth \rho \) \nnn
L_{1} &=& i e^{i u} \( \coth ( 2 \rho ) \p_u
- \frac{1}{ \sinh (2 \rho ) } \p_v
- \frac{i}{2} \p_\rho
+ \frac{i}{2} s \coth \rho \)
\eeq
where $u = \tau + \phi$ and $v = \tau - \phi$. 
There is also a barred set of generators ${\bar L}_0$, ${\bar L}_{-1}$
and ${\bar L}_1$ obtained through $u \leftrightarrow v$ and
$s \to -s$.  Thus the full subalgebra is $SL(2, \Rbf) \otimes SL(2, \Rbf)$.
In a Virasoro algebra, there is an infinite number of
generators $L_n$ with $n \in \Zbf$. 

We will now make some comments
that will make the role of the parameter $s$ clear.

In the conformal field theory, one seeks primary fields in order to
develop the modules associated with the Virasoro algebra.  These should
satisfy
\beq
&& L_1 \psi = {\bar L}_1 \psi = 0 \nnn
&& L_0 \psi = h \psi, \quad {\bar L}_0 \psi = {\bar h} \psi
\eeq
where $h$ and ${\bar h}$ are the conformal weights.  
($L_{-1}$ and ${\bar L}_{-1}$ behave as raising operators,
generating other parts of the spectrum.)
In order for
such primary fields to exist, one finds that $s = h - {\bar h}$,
which is a sort of ``spin.''  
The primary fields then take the form \cite{deBoer:1998kjm}:
\beq
\psi \sim \frac{e^{-i h u - i {\bar h} v}}{ ( \cosh \rho )^{ h + {\bar h}}}
= \frac{e^{-i (h + {\bar h}) \tau -i s \phi}}{ ( \cosh \rho )^{ h + {\bar h}}}
\label{ads3primary}
\eeq
From the last equation we can see why $s$ is called ``spin,'' seeing
how it is dual to the angular variable $\phi$.  On the other hand,
$h + {\bar h}$ is the total conformal dimension of the field.

What one finds from applying the corresponding Casimir operator of
$SL(2,\Rbf)$ built from $L_1$, $L_{-1}$ and $L_0$ is that the
primary field $\psi$ of spin $s$ satisfies the differential
equation \cite{deBoer:1998kjm}:
\beq
[ 2 h ( h - 1 ) + 2 {\bar h} ( {\bar h} - 1 ) ] \psi
= \ell^2 \Box \psi + s^2 \coth^2 \rho \psi
\label{casimir}
\eeq
This makes clear where the connection between the mass of the
field in the supergravity and the conformal weights $h$ and ${\bar h}$
of the boundary CFT comes from.  Note that there is a contribution
coming from the spin $s$ which calls for a nontrivial $\rho$ dependent
value of $\Box \psi$.  We also see some of the special features of 2d CFTs
as opposed to 4d CFTs, where both $h$ and ${\bar h}$ play a role,
including through the spin $s$, which is their difference.
This is closely related to the fact that in 2d one can
have right-movers and left-movers, which allows for a sort
of chirality even for bosonic fields.  Of course this is consistent
with the fact that in 2d we can perform bosonization of
fermionic modes.

To compute the ``Laplacian'' (or more accurately a D'Alembertian) associated
with $AdS_3$ in the coordinate system above, one has to work out $\nabla^2 \psi = 
\nabla^\mu \nabla_\mu \psi$ for a scalar field $\psi$
 in this nontrivial geometry.  Because of the curvature of
$AdS_3$, this entails working out
the Christoffel symbols so that the  covariant derivative
can be computed.  By straightforward calculations we find
\beq
&& \Gamma_{\tau \tau}^\tau = 0, \quad 
\Gamma_{\tau \rho}^\tau = \Gamma_{\rho \tau}^\tau = \tanh \rho, \quad
\Gamma_{\tau \phi}^\tau = \Gamma_{\phi \tau}^\tau = 0 \nnn
&& \Gamma_{\rho \rho}^\tau = 0, \quad
\Gamma_{\rho \phi}^\tau = \Gamma_{\phi \rho}^\tau = 0, \quad
\Gamma_{\phi \phi}^\tau = 0 \nnn
&& \Gamma_{\tau \tau}^\rho = \sinh \rho \cosh \rho, \quad
\Gamma_{\tau \rho}^\rho = \Gamma_{\rho \tau}^\rho = 0, \quad
\Gamma_{\tau \phi}^\rho = \Gamma_{\phi \tau}^\rho = 0 \nnn
&& \Gamma_{\phi \phi}^\rho = -\sinh \rho \cosh \rho, \quad
\Gamma_{\rho \rho}^\rho = 0, \quad
\Gamma_{\rho \phi}^\rho = \Gamma_{\phi \rho}^\rho = 0 \nnn
&& \Gamma_{\tau \tau}^\phi = 0, \quad
\Gamma_{\rho \rho}^\phi = 0, \quad
\Gamma_{\phi \phi}^\phi = 0 \nnn
&& \Gamma_{\tau \rho}^\phi = \Gamma_{\rho \tau}^\phi = 0, \quad
\Gamma_{\tau \phi}^\phi = \Gamma_{\phi \tau}^\phi = 0, 
\nnn
&&
\Gamma_{\rho \phi}^\phi = \Gamma_{\phi \rho}^\phi = \coth \rho
\eeq
Next, the D'Alembertian of a scalar field is given by
\beq
\nabla_\mu \nabla^\mu \psi &=& g^{\mu\nu} ( \p_\mu \p_\nu \psi - \Gamma_{\mu \nu}^\s \p_\s \psi )
\nnn
&=& g^{\tau \tau} ( \p_\tau^2 \psi - \Gamma_{\tau \tau}^\s \p_\s \psi )
\nnn
&& + g^{\rho \rho} ( \p_\rho^2 \psi - \Gamma_{\rho \rho}^\s \p_\s \psi )
\nnn
&& + g^{\phi \phi} ( \p_\phi^2 \psi - \Gamma_{\phi \phi}^\s \p_\s \psi )
\nnn
&=& -\frac{1}{\ell^2} \frac{1}{\cosh^2 \rho} \p_\tau^2 \psi
+ \frac{1}{\ell^2} \p_\rho^2 \psi
+ \frac{1}{\ell^2} \frac{1}{\sinh^2 \rho} \p_\phi^2 \psi
\nnn
&& + \frac{1}{\ell^2} ( \tanh \rho + \coth \rho ) \p_\rho \psi
\eeq
We can then apply this to the primary field \myref{ads3primary}.
We find:
\beq
\p_\tau^2 \psi &=& -(h + {\bar h})^2 \psi \nnn
\p_\phi^2 \psi &=& - s^2 \psi \nnn
\p_\rho \psi &=& -( h + {\bar h} ) \tanh \rho \cdot \psi \nnn
\p_\rho^2 \psi &=& -(h + {\bar h}) \sech^2 \rho \cdot \psi
+ (h + {\bar h})^2 \tanh^2 \rho \cdot \psi
\eeq
Then using these one finds after some hyperbolic trigonometric identities
that
\beq
\Box \psi &=& \nabla^\mu \nabla_\mu \psi
\nnn
&=& [ (h + {\bar h})^2 - 2 (h + {\bar h}) - s^2 \csch^2 \rho ]
\frac{\psi}{\ell^2}
\eeq
Now we perform the manipulation
\beq
(h + {\bar h})^2 - 2 (h + {\bar h})
= 2 h (h - 1) + 2 {\bar h} ( {\bar h} - 1) - s^2
\eeq
to obtain
\beq
\Box \psi =
[ 2 h (h - 1) + 2 {\bar h} ( {\bar h} - 1) - s^2 \coth^2 \rho ]
\frac{\psi}{\ell^2}
\eeq
It can be seen that this agrees with \myref{casimir} above.

The D1-D5 system can be described by 6d supergravity on $AdS_3 \times S^3$, or
the string version thereof.  Thus the low energy excitations of the 2d CFT
associated with D1-D5 can be extracted from a mode analysis of the 6d
gravity compactified on $S^3$.  For this, it is important to understand
representations of the isometry group $SO(4)$ of $S^3$ and how they
correspond to spherical harmonics on that space.  Supersymmetry is of
a big help here.

To get to a 6d SUGRA from string theory, one must compactify on a 4d manifold.
Two possibilities are considered in de Boer \cite{deBoer:1998kjm}:  $T^4$ and $K3$. Let's
start with K3.  For heterotic or
Type I, this gives a (0,1) theory.  For Type IIA, this gives a (1,1) theory.
For Type IIB this gives (0,2) theory.
In contrast, on $T^4$, Types IIA and IIB give a (2,2) theory.
Here the notation indicates left-handed and right-handed supersymmetries.
This is because in 2d or 10d, we can have Weyl-Majorana spinors.
Thus there are more possibilities than in 4d.  These are
denoted by $(n_L, n_R)$.

$AdS_3 \times S^3$ has isometry group $SO(2,2) \times SO(4)$
where the $SO(2,2)$ is associated with the $AdS_3$ and the $SO(4)$
is associated with $S^3$.  These can be seen by the coordinate descriptions
that embed these spaces in one higher dimension.  $AdS_3$ can be
described by the hyperbolic equation
\beq
X_1^2 + X_2^2 - X_3^2 - X_4^2 = R^2
\eeq
which has an obvious $SO(2,2)$ symmetry.  $S^3$ of course
can be described by
\beq
X_1^2 + X_2^2 + X_3^2 + X_4^2 = R^2
\eeq
which has an $SO(4)$ symmetry.  In terms of covering groups,
\beq
SO(2,2) \simeq SL(2,\Rbf) \otimes SL(2,\Rbf)
\eeq
and
\beq
SO(4) \simeq SU(2) \otimes SU(2)
\eeq
These are associated with left-handed and right-handed fields,
each of which fall into representations of $SL(2,\Rbf) \otimes SU(2)$.
These are actually embedded into larger groups $G_L$ and $G_R$, where
\beq
G_L \supset SL(2,\Rbf) \otimes SU(2)
\eeq
and similarly for $G_R$.  The related supergroup is $SU(1,1|2)$.
Thus we are particularly interested in the representation theory
of this supergroup, since we are dealing with a super-CFT.

There are some important analogies between the D1-D5 system
and $\Ncal=4$ super-Yang-Mills obtained through compactification
of superstring theory on $AdS_5 \times S^5$.  In the latter case,
only short multiplets of the superconformal algebra are involved
due to the 32 supersymmetries (4 Majorana supercharges with 4 components each
and 4 additional superconformal fermionic generators that are also Majorana
spinors with 4 components each) 
and the fact that the representations
must have at most spin 2 corresponding to the graviton.  In the
case of 6d supergravity with 8 or 16 supersymmetries, corresponding
to (1,0), (1,1) and (2,0), again only short multiplets are involved.
As shown by de Boer, this means that the KK spectrum and hence
states in the 2d CFT can be computed completely using representation
theory.

As usual with 2d CFTs with a large amount of supersymmetry, quite beautiful
structures arise once all charges are taken into account.  One has
$L_1$, $L_0$ and $L_{-1}$ defining an $SL(2,\Rbf)$ algebra.
Then one has $J_+$, $J_0$ and $J_-$ defining an $SU(2)$ algebra.
Finally, one has $Q^+$ and $Q^-$ that are supersymmetry generators
taking states back and forth from the bosonic side and fermionic
side.  Altogether this rounds out $SU(2|1,1)$.  One has this both
in the left movers and right movers in the case of (1,1) supersymmetry.
That gives an overall $SU(2|1,1) \otimes SU(2|1,1)$ superalgebra.
Or, in the case of (0,2) supersymmetry both $SU(2|1,1)$s are in the
right-moving sector.

We are interested in the CFT of the D1-D5 theory with the D5 branes
wrapped on a four-dimensional compact manifold.  A particular example
is when the D5 branes are wrapped on the well-known manifold K3.
In this case it has been conjectured that the CFT is a deformation
of a supersymmetric $\s$ model \cite{Vafa:1995zh,Gubser:1998bc}.  
Furthermore, it has been conjectured
that the target space of this deformed $\s$ model is
$K3^N/S_N$.  Here, $K3^N$ represents the product space built
from $N$ copies of K3.  Also, $S_N$ is the permutation group
of order $N$.  E.g., $S_3$ would contain the group operations
that give the 6 permutations
of $(1,2,3)$:
\beq
&& (1,2,3), \quad (2,3,1), \quad (3,1,2)
\ddd
(3,2,1), \quad (2,1,3), \quad (1,3,2)
\eeq
In the first line we give the 3 even permutations
and in the second line we give the 3 odd permutations.
For large $N$, $S_N$ contains a very large number
of elements.  The notation $K3^N/S_N$ indicates that
we have an equivalence between the points of the $K3^N$
space if they are related by the permutation group $S_N$.
For example, if $({\bf x}_1, {\bf x}_2, {\bf x}_3)$ is
a point in $K3 \times K3 \times K3$, with each
${\bf x}_i$ a four-component position vector, then
we would have the equivalence under the permutation $(2,1,3)$
of
\beq
({\bf x}_1, {\bf x}_2, {\bf x}_3)
\simeq ({\bf x}_2, {\bf x}_1, {\bf x}_3)
\eeq
This is an orbifold because it is a quotient space built using
a discrete group, $S_N$ (or $S_3$ in the specific example just
given).

In the analysis of the CFT, a key quantity is the Poincare polynomial \cite{deBoer:1998kjm}:
\beq
P_{t, {\bar t}} = \tr ( t^{J_0} {\bar t}^{{\bar J}_0} )
\eeq
Here, $J_0$ and ${\bar J}_0$ are the generators in the
Cartan subalgebra of the $SU(2) \otimes SU(2)$ internal
symmetry.  Also, the trace is only over the chiral
primaries of the CFT.  The quantities $t$ and ${\bar t}$
are independent complex numbers.
For instance, in counting states, de Boer was
particularly interested in arbitrary $t$
but ${\bar t}= -1$, and the case of the K3
manifold:  $P_{t, -1}(K3^N/S_N)$.

It turns out that this Poincare
polynomial is directly related to the manifold of the
target space in an interesting way:
\beq
P_{t, {\bar t}} = \sum_{p,q} h_{p,q} t^p {\bar t}^q
\eeq
Here, $h_{p,q}$ are the Betti numbers of the manifold.
Thus there is an intimate connection between homology
and this polynomial.

There is also an interesting connection here to orbifold CFT.
This is where we start with a CFT on a space $X$, and
then perform an orbifold of the target space,
\beq
X \to X^n / Z_n
\eeq
Or, if the CFT is $M$, then the orbifold is $M^n/Z_n$.
It turns out that if we understand the representations
of the conformal algebra on $X$, then there is a straightforward
relation to representations on $X^n / Z_n$.

The untwisted sector of the orbifold CFT has
\beq
h' = n h, \quad
q' = n q
\eeq
Here, $h$ is the conformal weight of the CFT $M$ and $q$ is the charge
under the internal symmetry.  The primed values correspond to the same
quantities in the orbifold CFT $M^n/Z_n$.

For the twisted sector labeled by $m$ (a non-negative integer), we have
\beq
h' = \frac{h+m}{n} + \frac{c}{24} \frac{n^2 - 1}{n},
\quad
q' = q
\eeq
where $c$ is the central charge of the CFT $M$.

Related to all of this is the generating function for the Poincare polynomial \cite{deBoer:1998kjm}:
\beq
\sum_{N \geq 0} Q^N P_{t, {\bar t}} ( M^N / S_N ) =
\prod_{m=1}^\infty \prod_{p,q} ( 1 + (-1)^{p+q+1} Q^m
t^{p + \frac{d}{2} (m-1)}
{\bar t}^{q + \frac{d}{2} (m-1)} )^{ (-1)^{p+q+1} h_{p,q} }
\eeq
One should think about $Q$ as being a small complex number.
Then one can take the $N$th derivative w.r.t.~$Q$ and set
$Q \to 0$ after differentiating.  On the l.h.s.~one then
obviously ends up with the quantity of interest,
$P_{t, {\bar t}} ( M^N / S_N )$.  Then on the r.h.s., one
obtains an expression for this in terms of the Betti numbers.

Through various calculations along these lines, de Boer is able
to count states, such as $67131 N - 244053$ states of the form
\beq
| q , h \rangle_L \otimes | q' , h' \rangle_R
= | 0 , 2 \rangle_L \otimes | q', q'/2 \rangle_R
\eeq
Thus we can learn various numerological data about orbifold
CFTs.  One could imagine tabulating such things almost
ad infinitum.
de Boer's motivation was to resolve certain quandries
about missing states noticed by Vafa \cite{Vafa:1998nt}.

\section{de Sitter holography}
Holographic ideas can also be applied to de Sitter spacetime and hence cosmology.  This
is described for instance in Skendaris et al.~\cite{Afshordi:2016dvb}.
(See also references 4-8 of that paper.) 
This particular approach has been coined ``holographic cosmology.''
It can be thought of as a holographic framework to address some
of the shortcomings of conventional inflationary theory.  Additionally,
it is motivated by the fact that quantum gravity appears to be
holographic in nature and there is little doubt that quantum gravity
also plays a role in the earliest stages of the universe.

This scenario has a number of interesting features:
\ben
\item The dual theory to the cosmological universe is a 3d quantum field theory (QFT).
\item This QFT is located at future infinity.
\item The partition function, in the presence of sources for gauge invariant
operators, is the wavefunction of the universe.
\item Cosmic evolution is mapped to (inverse) RG flow in the QFT.
\item It provides a model for a non-geometric early universe.
\item The fields in the QFT are taken to be in the adjoint representation.
Thus 3d super-Yang-Mills falls into the class of theories to be considered,
and may have a special cosmology.
\een

Standard $\Lambda$CDM has six adjustable parameters.  Skendaris et al.~holographic cosmology
addresses two of them:  the tilt $n_s$ and the scalar perturbation
power spectrum normalization.  In the conventional
cosmology these appear through the power spectrum:
\beq
\Delta^2_{\cal R}(q) = \Delta^2_0(q_*) \( \frac{q}{q_*} \)^{n_s-1}
\eeq
In the holographic cosmology there is a different prediction for the power spectrum, which is
given by
\beq
\Delta^2_{\cal R}(q) = \frac{\Delta^2_0(q_*)}{1 + (g q_*/q) \ln | q/\beta g q_* | + \ord{gq_*/q}^2} 
\eeq
In the discussion of these authors, one finds that they can no longer
trust their formula for small $\ell$ (the multipole moment of the CMB),
 because the two-loop term becomes competitive with the
one-loop term.  So really to handle the small $\ell$ behavior one needs to do a numerical simulation
and compute the two-point function of the energy momentum tensor at small $q$.
The analytic continuation (14) in their paper raises the question of how this is related
to an imaginary time correlator.
Of course to carry out a computation using lattice field
theory, one has to derive the energy momentum tensor.
This can be nontrivial.  For instance, in the formulation of ${\cal N}=4$ 3d super-Yang-Mills
on the lattice using the Blau-Thompson twist, it is a significant challenge
to find the combination of operators and renormalization constants
that really give a conserved energy-momentum tensor.  
This is something that we hope to complete in the future.
Then we would have a supersymmetric holographic cosmology.

In \cite{Nastase:2019rsn} it is shown how the standard
cosmological problems that motivate inflation are resolved in the
context of holographic cosmology.  More specifically, they look
at the horizon problem, the flatness problem and the problem of
relics (such as GUT monopoles).

For instance, for the horizon problem, points on the surface of
last scattering are correlated through correlations in the dual QFT
assuming they run deep enough into the IR.  While these super-renormalizable theories
are perturbatively IR divergent, it is believed that they are non-perturbatively
IR finite.  This then corresponds to a resolution of the cosmic initial
singularity.  This is certainly an interesting possibility.  As is well-known,
the presence of singularities in classical gravity has long intrigued physicists.
It has always been hoped that a theory of quantum gravity would somehow resolve
them.  In the present context, this is accomplished through the dual
QFT description, exploiting holography.

\section{SUSY warped extra dimension}
Now we briefly discuss some aspects of the 
low-energy effective field theory of formulating
physics on a slice of $AdS_5$.  The leading terms of the action are of course
\beq
S = -\int d^4 x dy \sqrt{-g} \bigg\{
\tr F^{MN} F_{MN}
+ D^M \phi^\dagger D_M \phi
+ \overline{\Psi} \Gamma^M D_M \Psi
\bigg\}
\eeq
That is, a gauge action, a scalar action and
a fermion action.
The indices will be chosen as $M = 0,1,\ldots,3$ or $5$, where $M=5$ denotes
the extra dimension; i.e., corresponding to the coordinate $x^5=y$ in \myref{ymetric} below.  
The metric appears implicitly in a few places, such
as $F^{MN} = g^{MP} g^{NQ} F_{PQ}$, $D^M = g^{MN} D_N$ and $\Gamma^M = e^M_A \Gamma^A$.
Here, $\{ \Gamma^A, \Gamma^B \} = -2 \eta^{AB}$ with $\eta^{AB} =
\text{diag} (-1, 1, 1, 1, 1)$.  As usual, the funfbein is related to the
metric, such as $g^{MN} = e^M_A e^N_B \eta^{AB}$.

So far, our description is general to any five-dimensional (5d) spacetime.
We now want to specify the Randall-Sundrum geometry, which is a slice
of 5d anti-deSitter (AdS$_5$) space.  There are several coordinate
systems that we could choose from.  Here we highlight one of the more
common coordinate systems.
The 5d metric is given by
\beq
ds^2 = e^{-2k|y|} (-dt^2 + d{\bf x}^2 ) + dy^2
\label{ymetric}
\eeq
where $k$ is related to the inverse of the AdS radius $L$ and is generally
of order the 4d Planck mass.  The extra dimension $y$ is taken to be
an orbifold, $S^1/Z_2$.  That is, one begins with a circle, $y \simeq
y + 2\pi R$.  Then a further identification is made:  $y \simeq -y$.
This results in two fixed points $y=0$ and $y=\pi R$.  Points 
$y \in (\pi R, 2 \pi R)$ are identified with points $y \in (0, \pi R)$
since on $S^1/Z^2$ we have $y \simeq -y \simeq 2 \pi R - y$.

Next one performs a KK decomposition on the space $S^1/Z_2$.  To complete
this process, one must know the orbifold action on the fields.  In
particular, the fermions are subject to even or odd conditions depending
on their chirality. 

There is another coordinate system where the
extra dimension is denoted as $z$, with
metric
\beq
ds^2 = \frac{L^2}{z^2} (-dt^2 + d{\bf x}^2 + dz^2 )
\label{zmetric}
\eeq
We wish to relate this to the metric \myref{ymetric} above.  
By comparing these two we can figure out the change of
coordinates.  For instance matching the $dt^2$ terms we see that
\beq
e^{-2k|y|} = \frac{L^2}{z^2}
\eeq
Taking the $y > 0$ branch, we have that
\beq
z = L e^{ky}, \quad \frac{dz}{z} = k \, dy
\eeq
or
\beq
\frac{L^2}{z^2} dz^2 = (kL)^2 dy^2
\eeq
Finally, to match the two metrics, we have
\beq
kL = 1
\eeq
so that as stated, $k$ is the inverse of the AdS$_5$ radius $L$.
It will turn out that \myref{zmetric} is convenient for generalizations
of AdS$_5$ that are inspired by developments in string theory.

\section{Deformed AdS}
In \cite{Gabella:2007cp} we considered a deformation of $AdS_5$ that
was inspired by some supergravity constructions that broke supersymmetry \cite{Kuperstein:2003yt}.
The idea was that by doing so we would be able to have supersymmetry
breaking arise from the background geometry, but now in a Randall-Sundrum
type phenomenological theory.
 
The geometry is
\beq
ds^2 &=& A^2(z) ( -dt^2 + d{\bf x}^2 + dz^2 ) \nnn
A^2(z) &=& \frac{1}{(kz)^2} \[ 1 - \e \( \frac{z}{z_1} \)^4 \]
\eeq
with $z_0 \leq z \leq z_1$.  $z_0$ is the position of the UV brane in the
extra dimension.  $z_1$ is the position of the IR brane.  If $\e=0$
we would have a slice of $AdS_5$.  The presence of the term with
coefficient $\e$ is supposed to be dual to dynamical supersymmetry
breaking in the supersymmetric gauge theory.

The fermion equation of motion (free particle limit) after the rescaling
${\hat \Psi} = A^2 \Psi$ is
\beq
( \delta_\alpha^\mu \gamma^\alpha \p_\mu + \gamma_5 \p_z + cA ) {\hat \Psi} = 0
\eeq
In the slice of AdS, the scalar mass is given by:
\beq
M_\Phi^2 = a k^2 + 2 b k^2 z [ \delta(z - z_0) - \delta(z - z_1) ]
\eeq
Thus the parameters $a$ and $b$ are key to the construction. 
The equation of motion for the scalars is:
\beq
\p_\mu \p^\mu \Phi + A^{-3} \p_5 ( A^3 \p^5 \Phi )
- a k^2 A^2 \Phi = 0
\eeq
This equation is solved by separation of variables:
\beq
\Phi(x,z) = \sum_{n=0}^\infty \phi_n(x) {\tilde f}_n(z)
\eeq
The functions ${\tilde f}_n(z)$ are known as the ``profiles'' of the Kaluza-Klein (KK)
modes.  In the supersymmetric limit, which corresponds to $\e \to 0$ in the
deformed metric, zeromodes are produced when the modified Neumann conditions
\beq
\left. ( {\tilde f}_n^{\prime} - b k^2 z A^2 {\tilde f}_n ) \right|_{z_0, z_1} = 0
\eeq
are imposed.  This also require that $b$ and $a$ be related by
\beq
b = 2 \pm \sqrt{ 4 + a }
\eeq
In the $\e \not= 0$ case, deviation from this produces masses for the scalars,
leading to a supersymmetry violating splitting in the chiral multiplets.

\section{Conclusions}
We have seen that there are many interesting examples of holography beyond
the standard $AdS_5 \times S^5$ construction.  In some cases they have been
brought quite close to phenomenology.  This can occur through reducing the
amount of supersymmetry or eliminating conformal invariance.  It can also
consist of a bottom-up approach that is inspired by string theory and supergravity.
The D1-D5 system also shows that holographic duality for 2d CFTs and $AdS_3$
can be rich in representation theory, generating functions, Virasoro
operators and other nice mathematical features.  We described some of
the difficulties of formulating these theories in terms of a lattice
theory that can be simulated.  However, we believe that with consistent
effort progress can also be made in that direction.  Then one would be
able to compute features of quantum gravity by simulating a QFT.
A interesting example of this occurs in holographic cosmology, which
we briefly described.

\section*{Acknowledgements}
The author was supported in part by the Department of Energy, Office of Science, Office of High Energy Physics,
Grant No.~DE-SC0013496.

\bibliography{kwd7_rvw}

\providecommand{\href}[2]{#2}\begingroup\raggedright\begin{thebibliography}{10}

\bibitem{Susskind:1978ms}
L.~Susskind, \emph{{Dynamics of Spontaneous Symmetry Breaking in the Weinberg-
  Salam Theory}}, \href{https://doi.org/10.1103/PhysRevD.20.2619}{\emph{Phys.
  Rev.} {\bfseries D20} (1979) 2619}.

\bibitem{Weinberg:1979bn}
S.~Weinberg, \emph{{Implications of Dynamical Symmetry Breaking: An Addendum}},
  \href{https://doi.org/10.1103/PhysRevD.19.1277}{\emph{Phys. Rev.} {\bfseries
  D19} (1979) 1277}.

\bibitem{Holdom:1981rm}
B.~Holdom, \emph{{Raising the Sideways Scale}},
  \href{https://doi.org/10.1103/PhysRevD.24.1441}{\emph{Phys. Rev.} {\bfseries
  D24} (1981) 1441}.

\bibitem{Holdom:1984sk}
B.~Holdom, \emph{{Techniodor}},
  \href{https://doi.org/10.1016/0370-2693(85)91015-9}{\emph{Phys. Lett.}
  {\bfseries B150} (1985) 301}.

\bibitem{Yamawaki:1985zg}
K.~Yamawaki, M.~Bando and K.-i. Matumoto, \emph{{Scale Invariant Technicolor
  Model and a Technidilaton}},
  \href{https://doi.org/10.1103/PhysRevLett.56.1335}{\emph{Phys. Rev. Lett.}
  {\bfseries 56} (1986) 1335}.

\bibitem{Bando:1986bg}
{M.~Bando, K.~Matumoto, and K.~Yamawaki}, \emph{{Technidilaton}},
  \href{https://doi.org/10.1016/0370-2693(86)91516-9}{\emph{Phys. Lett.}
  {\bfseries B178} (1986) 308}.

\bibitem{Appelquist:1986an}
T.~W. Appelquist, D.~Karabali and L.~C.~R. Wijewardhana, \emph{{Chiral
  Hierarchies and the Flavor Changing Neutral Current Problem in Technicolor}},
  \href{https://doi.org/10.1103/PhysRevLett.57.957}{\emph{Phys. Rev. Lett.}
  {\bfseries 57} (1986) 957}.

\bibitem{Appelquist:1986tr}
T.~Appelquist and L.~C.~R. Wijewardhana, \emph{{Chiral Hierarchies and Chiral
  Perturbations in Technicolor}},
  \href{https://doi.org/10.1103/PhysRevD.35.774}{\emph{Phys. Rev.} {\bfseries
  D35} (1987) 774}.

\bibitem{Appelquist:1987fc}
T.~Appelquist and L.~C.~R. Wijewardhana, \emph{{Chiral Hierarchies from Slowly
  Running Couplings in Technicolor Theories}},
  \href{https://doi.org/10.1103/PhysRevD.36.568}{\emph{Phys. Rev.} {\bfseries
  D36} (1987) 568}.

\bibitem{Appelquist:1993sg}
T.~Appelquist and J.~Terning, \emph{{An Extended technicolor model}},
  \href{https://doi.org/10.1103/PhysRevD.50.2116}{\emph{Phys. Rev. D}
  {\bfseries 50} (1994) 2116}
  [\href{https://arxiv.org/abs/hep-ph/9311320}{{\ttfamily hep-ph/9311320}}].

\bibitem{Kribs:2016cew}
G.~D. Kribs and E.~T. Neil, \emph{{Review of strongly-coupled composite dark
  matter models and lattice simulations}},
  \href{https://doi.org/10.1142/S0217751X16430041}{\emph{Int. J. Mod. Phys. A}
  {\bfseries 31} (2016) 1643004}
  [\href{https://arxiv.org/abs/1604.04627}{{\ttfamily 1604.04627}}].

\bibitem{DeGrand:2019vbx}
T.~DeGrand and E.~T. Neil, \emph{{Repurposing lattice QCD results for composite
  phenomenology}},
  \href{https://doi.org/10.1103/PhysRevD.101.034504}{\emph{Phys. Rev. D}
  {\bfseries 101} (2020) 034504}
  [\href{https://arxiv.org/abs/1910.08561}{{\ttfamily 1910.08561}}].

\bibitem{Maldacena:1997re}
J.~M. Maldacena, \emph{{The Large N limit of superconformal field theories and
  supergravity}}, \href{https://doi.org/10.1023/A:1026654312961}{\emph{Adv.
  Theor. Math. Phys.} {\bfseries 2} (1998) 231}
  [\href{https://arxiv.org/abs/hep-th/9711200}{{\ttfamily hep-th/9711200}}].

\bibitem{Klebanov:2000hb}
I.~R. Klebanov and M.~J. Strassler, \emph{{Supergravity and a confining gauge
  theory: Duality cascades and chi SB resolution of naked singularities}},
  \href{https://doi.org/10.1088/1126-6708/2000/08/052}{\emph{JHEP} {\bfseries
  08} (2000) 052} [\href{https://arxiv.org/abs/hep-th/0007191}{{\ttfamily
  hep-th/0007191}}].

\bibitem{Afshordi:2016dvb}
N.~Afshordi, C.~Coriano, L.~Delle~Rose, E.~Gould and K.~Skenderis, \emph{{From
  Planck data to Planck era: Observational tests of Holographic Cosmology}},
  \href{https://doi.org/10.1103/PhysRevLett.118.041301}{\emph{Phys. Rev. Lett.}
  {\bfseries 118} (2017) 041301}
  [\href{https://arxiv.org/abs/1607.04878}{{\ttfamily 1607.04878}}].

\bibitem{deBoer:1998kjm}
J.~de~Boer, \emph{{Six-dimensional supergravity on S**3 x AdS(3) and 2-D
  conformal field theory}},
  \href{https://doi.org/10.1016/S0550-3213(99)00160-1}{\emph{Nucl. Phys. B}
  {\bfseries 548} (1999) 139}
  [\href{https://arxiv.org/abs/hep-th/9806104}{{\ttfamily hep-th/9806104}}].

\bibitem{Romans:1984an}
L.~J. Romans, \emph{{New Compactifications of Chiral $N=2 d=10$ Supergravity}},
  \href{https://doi.org/10.1016/0370-2693(85)90479-4}{\emph{Phys. Lett. B}
  {\bfseries 153} (1985) 392}.

\bibitem{Klebanov:1998hh}
I.~R. Klebanov and E.~Witten, \emph{{Superconformal field theory on
  three-branes at a Calabi-Yau singularity}},
  \href{https://doi.org/10.1016/S0550-3213(98)00654-3}{\emph{Nucl. Phys. B}
  {\bfseries 536} (1998) 199}
  [\href{https://arxiv.org/abs/hep-th/9807080}{{\ttfamily hep-th/9807080}}].

\bibitem{Aharony:1998xz}
O.~Aharony, A.~Fayyazuddin and J.~M. Maldacena, \emph{{The Large N limit of
  N=2, N=1 field theories from three-branes in F theory}},
  \href{https://doi.org/10.1088/1126-6708/1998/07/013}{\emph{JHEP} {\bfseries
  07} (1998) 013} [\href{https://arxiv.org/abs/hep-th/9806159}{{\ttfamily
  hep-th/9806159}}].

\bibitem{Levi:2005hh}
T.~S. Levi and P.~Ouyang, \emph{{Mesons and flavor on the conifold}},
  \href{https://doi.org/10.1103/PhysRevD.76.105022}{\emph{Phys. Rev. D}
  {\bfseries 76} (2007) 105022}
  [\href{https://arxiv.org/abs/hep-th/0506021}{{\ttfamily hep-th/0506021}}].

\bibitem{Gherghetta:2006yq}
T.~Gherghetta and J.~Giedt, \emph{{Bulk fields in AdS(5) from probe D7
  branes}}, \href{https://doi.org/10.1103/PhysRevD.74.066007}{\emph{Phys. Rev.}
  {\bfseries D74} (2006) 066007}
  [\href{https://arxiv.org/abs/hep-th/0605212}{{\ttfamily hep-th/0605212}}].

\bibitem{Gabella:2007cp}
M.~Gabella, T.~Gherghetta and J.~Giedt, \emph{{A Gravity dual and LHC study of
  single-sector supersymmetry breaking}},
  \href{https://doi.org/10.1103/PhysRevD.76.055001}{\emph{Phys. Rev.}
  {\bfseries D76} (2007) 055001}
  [\href{https://arxiv.org/abs/0704.3571}{{\ttfamily 0704.3571}}].

\bibitem{Randall:1999ee}
L.~Randall and R.~Sundrum, \emph{{A Large mass hierarchy from a small extra
  dimension}}, \href{https://doi.org/10.1103/PhysRevLett.83.3370}{\emph{Phys.
  Rev. Lett.} {\bfseries 83} (1999) 3370}
  [\href{https://arxiv.org/abs/hep-ph/9905221}{{\ttfamily hep-ph/9905221}}].

\bibitem{Giddings:2001yu}
S.~B. Giddings, S.~Kachru and J.~Polchinski, \emph{{Hierarchies from fluxes in
  string compactifications}},
  \href{https://doi.org/10.1103/PhysRevD.66.106006}{\emph{Phys. Rev.}
  {\bfseries D66} (2002) 106006}
  [\href{https://arxiv.org/abs/hep-th/0105097}{{\ttfamily hep-th/0105097}}].

\bibitem{Goldberger:1999uk}
W.~D. Goldberger and M.~B. Wise, \emph{{Modulus stabilization with bulk
  fields}}, \href{https://doi.org/10.1103/PhysRevLett.83.4922}{\emph{Phys. Rev.
  Lett.} {\bfseries 83} (1999) 4922}
  [\href{https://arxiv.org/abs/hep-ph/9907447}{{\ttfamily hep-ph/9907447}}].

\bibitem{Strassler:2005qs}
M.~J. Strassler, \emph{{The Duality cascade}},  in \emph{{Theoretical Advanced
  Study Institute in Elementary Particle Physics (TASI 2003): Recent Trends in
  String Theory}}, pp.~419--510, 5, 2005,
  \href{https://doi.org/10.1142/9789812775108_0005}{DOI}
  [\href{https://arxiv.org/abs/hep-th/0505153}{{\ttfamily hep-th/0505153}}].

\bibitem{Gwyn:2007qf}
R.~Gwyn and A.~Knauf, \emph{{Conifolds and geometric transitions}},
  \href{https://doi.org/10.1103/RevModPhys.80.1419}{\emph{Rev. Mod. Phys.}
  {\bfseries 8012} (2008) 1419}
  [\href{https://arxiv.org/abs/hep-th/0703289}{{\ttfamily hep-th/0703289}}].

\bibitem{Seiberg:1994pq}
N.~Seiberg, \emph{{Electric - magnetic duality in supersymmetric nonAbelian
  gauge theories}},
  \href{https://doi.org/10.1016/0550-3213(94)00023-8}{\emph{Nucl. Phys. B}
  {\bfseries 435} (1995) 129}
  [\href{https://arxiv.org/abs/hep-th/9411149}{{\ttfamily hep-th/9411149}}].

\bibitem{Gopakumar:1998ki}
R.~Gopakumar and C.~Vafa, \emph{{On the gauge theory / geometry
  correspondence}},
  \href{https://doi.org/10.4310/ATMP.1999.v3.n5.a5}{\emph{Adv. Theor. Math.
  Phys.} {\bfseries 3} (1999) 1415}
  [\href{https://arxiv.org/abs/hep-th/9811131}{{\ttfamily hep-th/9811131}}].

\bibitem{Elituv:2018byg}
J.~Elituv, \emph{{A companion to "Non-K\"ahler Deformed Conifold, Ultra-Violet
  Completion and Supersymmetric Constraints in the Baryonic Branch"}},
  \href{https://arxiv.org/abs/1806.01840}{{\ttfamily 1806.01840}}.

\bibitem{Costello:2020jbh}
K.~Costello and N.~M. Paquette, \emph{{Twisted Supergravity and Koszul Duality:
  A case study in AdS$_3$}},
  \href{https://arxiv.org/abs/2001.02177}{{\ttfamily 2001.02177}}.

\bibitem{Giedt:2011zza}
J.~Giedt, \emph{{A deconstruction lattice description of the D1/D5 brane
  world-volume gauge theory}},
  \href{https://doi.org/10.1155/2011/241419}{\emph{Adv. High Energy Phys.}
  {\bfseries 2011} (2011) 241419}.

\bibitem{Binetruy:2000zx}
P.~Binetruy, G.~Girardi and R.~Grimm, \emph{{Supergravity couplings: A
  Geometric formulation}},
  \href{https://doi.org/10.1016/S0370-1573(00)00085-5}{\emph{Phys. Rept.}
  {\bfseries 343} (2001) 255}
  [\href{https://arxiv.org/abs/hep-th/0005225}{{\ttfamily hep-th/0005225}}].

\bibitem{Eloy:2021fhc}
C.~Eloy, G.~Larios and H.~Samtleben, \emph{{Triality and the consistent
  reductions on AdS$_{3}$ \texttimes{} S$^{3}$}},
  \href{https://doi.org/10.1007/JHEP01(2022)055}{\emph{JHEP} {\bfseries 01}
  (2022) 055} [\href{https://arxiv.org/abs/2111.01167}{{\ttfamily
  2111.01167}}].

\bibitem{Vafa:1995zh}
C.~Vafa, \emph{{Gas of d-branes and Hagedorn density of BPS states}},
  \href{https://doi.org/10.1016/0550-3213(96)00025-9}{\emph{Nucl. Phys. B}
  {\bfseries 463} (1996) 415}
  [\href{https://arxiv.org/abs/hep-th/9511088}{{\ttfamily hep-th/9511088}}].

\bibitem{Gubser:1998bc}
S.~S. Gubser, I.~R. Klebanov and A.~M. Polyakov, \emph{{Gauge theory
  correlators from noncritical string theory}},
  \href{https://doi.org/10.1016/S0370-2693(98)00377-3}{\emph{Phys. Lett. B}
  {\bfseries 428} (1998) 105}
  [\href{https://arxiv.org/abs/hep-th/9802109}{{\ttfamily hep-th/9802109}}].

\bibitem{Vafa:1998nt}
C.~Vafa, \emph{{Puzzles at large N}},
  \href{https://arxiv.org/abs/hep-th/9804172}{{\ttfamily hep-th/9804172}}.

\bibitem{Nastase:2019rsn}
H.~Nastase and K.~Skenderis, \emph{{Holography for the very early Universe and
  the classic puzzles of Hot Big Bang cosmology}},
  \href{https://doi.org/10.1103/PhysRevD.101.021901}{\emph{Phys. Rev. D}
  {\bfseries 101} (2020) 021901}
  [\href{https://arxiv.org/abs/1904.05821}{{\ttfamily 1904.05821}}].

\bibitem{Kuperstein:2003yt}
S.~Kuperstein and J.~Sonnenschein, \emph{{Analytic nonsupersymmtric background
  dual of a confining gauge theory and the corresponding plane wave theory of
  hadrons}}, \href{https://doi.org/10.1088/1126-6708/2004/02/015}{\emph{JHEP}
  {\bfseries 02} (2004) 015}
  [\href{https://arxiv.org/abs/hep-th/0309011}{{\ttfamily hep-th/0309011}}].

\end{thebibliography}\endgroup
\bibliographystyle{JHEP}

\end{document}